\documentclass[12pt]{article}
\usepackage{epsfig}
\usepackage{graphicx}
\usepackage{natbib}
\usepackage{epstopdf}
\usepackage{amsmath}
\usepackage{subfigure}
\usepackage{gensymb}
\begin{document}

\title{Experimental study of internal wave generation by convection in water}

\maketitle

\author{Michael~Le~Bars$^{1,2}$(lebars@irphe.univ-mrs.fr), Daniel~Lecoanet$^{3,4}$, St\'ephane~Perrard$^{1}$, Adolfo~Ribeiro$^{2}$, Laetitia~Rodet$^{1}$, Jonathan~M.~Aurnou$^{2}$, Patrice~Le~Gal$^{1}$
\\
\\

{$^1$CNRS, Aix-Marseille Universit\'e, Ecole Centrale Marseille, IRPHE UMR 7342, 49 rue F. Joliot-Curie, 13013 Marseille, France}\\

{$^2$Department of Earth, Planetary, and Space Sciences, University of California, Los Angeles, CA 90095-1567, USA}\\

{$^3$Department of Astrophysics and Theoretical Astrophysics, University of California, Berkeley, CA 94720, USA}\\

{$^4$Kavli Institute for Theoretical Physics, University of California, Santa Barbara, CA 93106, USA}
}

\begin{abstract}
We experimentally investigate the dynamics of water cooled from below at $0\degree {\rm C}$ and heated from above. Taking advantage of the unusual property that water's density maximum is at about $4\degree {\rm C}$, this set-up allows us to simulate in the laboratory a turbulent convective layer adjacent to a stably stratified layer, which is representative of atmospheric and stellar conditions. High precision temperature and velocity measurements are described, with a special focus on the convectively excited internal waves propagating in the stratified zone. Most of the convective energy is at low frequency, and corresponding waves are localized to the vicinity of the interface. However, we show that some energy radiates far from the interface, carried by shorter horizontal wavelength, higher frequency waves. Our data suggest that the internal wave field is passively excited by the convective fluctuations, and the wave propagation is correctly described by the dissipative linear wave theory.
\end{abstract}

\vspace{2pc}
\noindent{\it Keywords}: internal waves, stratified flows, Rayleigh-B\'enard convection

%%%%%%%%%%%%%%%%%%%%%%%%%%%%%%%%%%%%%%%%%%%%%%%%%%
%%%%%%%%%%%%%%%%%%%%%%%%%%%%%%%%%%%%%%%%%%%%%%%%%%

\section{Introduction}

%%%%%%%%%%%%%% Outline of the paper %%%%%%%%%%%%%%%%%
In many geophysical and astrophysical systems, a turbulent convective fluid layer is separated from a stably stratified layer by a relatively sharp but deformable interface. Examples include convective and radiative zones in stars, the atmospheric convective layer and overlying stratosphere, the Earth's oceans, and possibly the Earth's outer core according to the latest estimates of its thermal conductivity \cite[][]{hirose2013composition}. While motions in the stratified layers are often neglected, these layers actually support oscillatory motions called internal waves, which can be excited by the adjacent convection. The remote observation of these internal waves can be used to probe the interior of stars via asteroseismology \cite[e.g.][]{garcia2007tracking}. In addition, internal waves transport energy and momentum \cite[e.g.][]{bretherton1969mean, zahn1997angular,rogers2013internal}. They interact non-linearly and produce mean motions, responsible for, e.g., the Quasi-Biennial Oscillation of the equatorial zonal wind in the Earth's tropical stratosphere \cite[][]{plumb1978instability}, and possibly explaining the observed misalignment between the orbits of extrasolar planets and their host stars \cite[][]{rogers2012internal}. Internal waves can also break, mixing the ambient fluid.  This could have primordial consequences for stellar internal composition evolution \cite[][]{charbonnel2005influence} and result in mass loss during the late stages of massive stars evolution \cite[][]{quataert2012wave}. Internal waves are thus essential for accurate models of global climate and stellar dynamics. 

Global integrated models require length scales, time scales and velocity scales spanning many orders of magnitude to simultaneously resolve motions in turbulent convective and stratified zones, and to understand the details of the highly non-linear couplings between convective turbulence, waves and meridional circulation. This is clearly very challenging both analytically and numerically. Recent studies have estimated analytically the flux of internal waves \cite[e.g.][]{lecoanet2013internal} and their influence on the angular velocity profile \cite[e.g.][]{alvan2013coupling}, using models of convective turbulence and assumptions about the excitation mechanism. The relevance of these models remain to be validated. In particular, in both atmospheric \cite[e.g.,][]{ansong2010internal} and astrophysics \cite[e.g.,][]{Goldreich1990} communities, the question remains: does wave excitation mainly come from interface deflections or from bulk stresses within the convective zone?
The exponential increase in computing resources has recently allowed for the first three-dimensional nonlinear numerical simulations of convective turbulence and internal waves simultaneously \cite[][]{brun2011modeling, alvan2014theoretical}, but only at moderate turbulent intensity. Clearly, it would be computationally much more efficient to simulate the convective turbulence and the internal waves separately using dedicated numerical schemes \cite[][]{belkacem2009stochastic}. Crucially, this would necessitate a comprehensive description of the coupling between the two zones. In this context, quantitative data from a well-controlled experiment, in a simplified but physically motivated configuration, is of great interest to validate analytic theories and numerical simulations. 

Several experimental studies of internal wave generation by convective turbulence have been published since the 60's. Most studies focused on the transient case where an initially thermally stratified layer of water is suddenly heated from below \cite[e.g.][]{deardorff1969laboratory}: a convective mixed zone then forms at the bottom of the tank and progressively invades the whole depth of the fluid. This configuration was in particular investigated by \cite{michaelian2002coupling} using modern techniques of flow measurements. Their focus was mostly on the evolution of the interface position and on the fluctuations of the heat flux.  Furthermore, this work provides the first, and---to the best of our knowledge---the only experimental velocity maps of internal waves excited by convection, which were measured using the Correlation Image Velocimetry (CIV) technique. They observed intermittence in the internal wave motions, dominated by two types of flows: first, short wavelength, high frequency waves with velocity vectors with inclinations of $\theta = 55\degree \pm 5\degree$ from the horizontal; and later, long wavelength, low frequency waves corresponding to almost horizontal motions. 
Using the dispersion relation for internal waves,
\begin{equation}
\omega= N \sin{\theta}, \label{dispersion}
\end{equation}
where $\omega$ is the wave frequency, $N$ the buoyancy frequency and $\theta$ the angle between the wave vector and the vertical (equivalently, the angle between the isophase lines or the velocity vectors direction and the horizontal), \cite{michaelian2002coupling} interpreted these results as follows. High frequency waves, whose propagation angle is surprisingly constant over the range of stratifications studied, are excited by intermittent plumes which emanate from along the bottom boundary, especially early in the experiment. Later, a large-scale circulation develops, which excites low frequency waves by organizing the previously random plume motions.

Other experimental studies used salted water as a working fluid, focusing on the effects of an isolated buoyant element impacting a stratified layer. For instance, \cite{mclaren1973investigation} and \cite{cerasoli1978experiments} investigated the nature and energy transfer of internal gravity waves generated by a thermal (i.e. the instantaneous release of a fixed volume of buoyant fluid) using a particle tracking method and local conductivity  measurements plus dye line visualization. More recently, \cite{ansong2010internal} experimentally studied the generation of internal gravity waves by a turbulent plume (i.e., the continuous release of buoyant fluid) impinging upon the interface between a uniform density layer of fluid and a linearly stratified layer. Using non-intrusive axisymmetric synthetic Schlieren measurements, they determined the fraction of the energy flux carried by the waves. They also noticed that the dominant wave frequency lies in a narrow range relative to the buoyancy frequency, with $\theta \in [27\degree - 58\degree]$ and a mean value of around $45\degree$. Following \cite{dohan2003internal}, they hypothesized that turbulent fluctuations interact resonantly with the excited internal waves to optimize the vertical transport of horizontal
momentum, which for a fixed displacement amplitude takes place at $\theta = 45\degree$.

A particularly original experiment was performed by \cite{townsend1964natural}, using the unusual property that water's density maximum is at about $4\degree {\rm C}$. Consequently, in a tank filled with water with an horizontal bottom surface at $0\degree {\rm C}$ and a hotter upper surface, the density stratification is unstable at temperatures below $4\degree {\rm C}$ and stable at temperatures above. In a sufficiently tall tank, it is possible to produce a turbulent convective layer adjacent to a stably stratified layer in a coherent self-organizing system, which is representative of atmospheric and stellar conditions. \cite{townsend1964natural} used temperature measurements and dye visualization to detect internal waves in the stratified layer, but only found low frequency waves localised close to the interface. It is unclear whether there were waves propagating away from the interface along a ``favorite angle'', as described in the closely related configurations studied by \cite{michaelian2002coupling} and \cite{ansong2010internal}.  

In the present paper, we re-investigate this ``$4\degree {\rm C}$ experiment'' first introduced by \cite{townsend1964natural} using high precision local temperature measurements as well as global non-intrusive velocity measurements by particle image velocimetry (PIV), both in the convective and stratified zones. Our work builds on the preliminary results presented in the conference proceeding by \cite{perrard}. Our main objectives are to better describe the characteristics of the excited wave field and to provide quantitative data for validating analytical and numerical models. The paper is organized as follows. In section 2, we present the experimental set-up and the diagnostic tools. Section 3 is devoted to the statistics of the temperature fluctuations. The main results from PIV measurements are given in section 4. Finally, conclusions and open questions are addressed briefly in section 5. 
 \\

%%%%%%%%%%%%%%%%%%%%%%%%%%%%%%%%%%%%%%%%%%%%%%%%%%
%%%%%%%%%%%%%%%%%%%%%%%%%%%%%%%%%%%%%%%%%%%%%%%%%%
\section{Experimental set-up}

\begin{center}
\begin{figure}
\begin{center}\includegraphics[width=\linewidth]{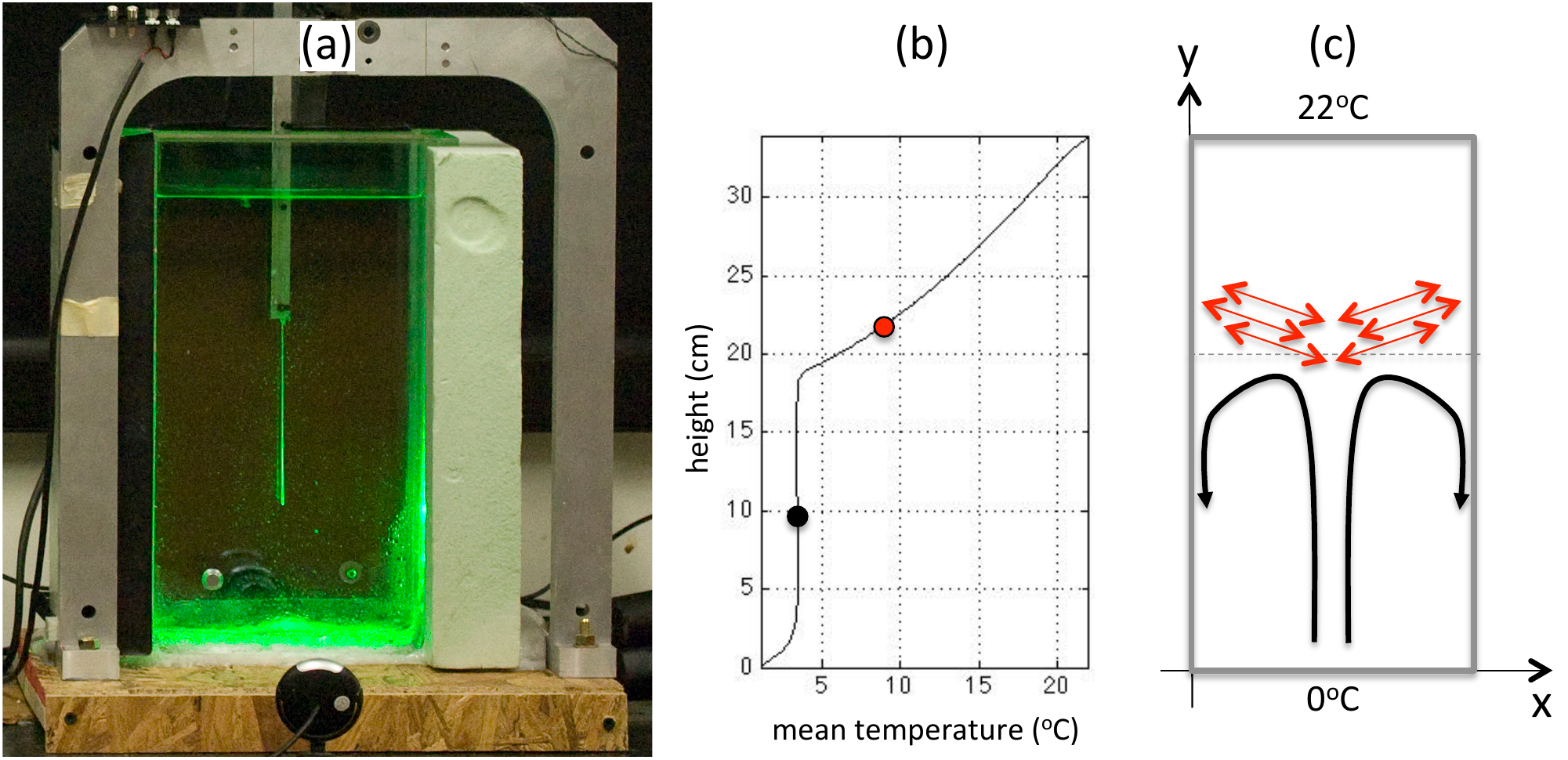}\end{center}
\caption{($a$) Picture of the experimental set-up, showing the tank, the two thermistor probes mounted on a vertical linear stage, and the green laser light sheet for PIV measurements. In this picture, the front and bottom foam insulating plates and the top copper plate have been removed for visualization. ($b$) Example of a time-averaged vertical temperature profile measured several hours after initializing an experiment. The black (red) dot shows schematically the location where the longterm temperature fluctuations are measured in the convective zone (stratified zone); see an example of temperature measurements in figure \ref{fig_data_T}. ($c$) Schematic of the expected flow with a large convective cell plus turbulent fluctuations in the lower convective zone and internal waves propagating from the interface into the upper stratified zone.}
\label{fig_setup}
\end{figure}
\end{center}

Our experimental set-up is presented in figure \ref{fig_setup}($a$). It consists of a rectangular tank whose lateral boundaries are made of vacuum glass along the large sides and $3 \ \text{cm}$ thick acrylic along the short sides. The internal dimensions are $18 \ \text{cm} \times 4 \ \text{cm}$ in width and $34 \ \text{cm}$ in height. Given the aspect ratios of our tank, no significative motion takes place in the short direction. However, based on the velocity measurements presented in section \ref{secVel}, the typical Reynolds number of our flow is about $110$ in the convective zone and $1.5$ in the stratified zone: the flow is thus influenced by viscous boundary layers from the side walls, especially in the stratified zone. The lower and upper boundaries are copper plates, whose temperatures are controlled within $\pm 0.05\degree {\rm C}$ by two circulating thermostated baths, with a fixed temperature equal to $0\degree {\rm C}$ and $22\degree {\rm C}$ respectively. To increase the thermal insulation of the tank, foam plates cover all boundaries, except when side view visualization is required, for instance during PIV measurements: foam plates from the two large sides are then removed, hence increasing heat losses. With all foam plates in place, the global external convective heat transfer coefficient of our set-up is about $2.9 \ {\rm W} \, {\rm m}^{-2} \, {\rm K}^{-1}$, which would correspond to $7 \ \text{cm}$ thick acrylic walls. At the beginning of each experiment, the baths are first allowed to reach their assigned temperature. The tank is then filled with water at about $4\degree {\rm C}$ up to a given depth, chosen according to estimates of heat flux equilibrium or according to previous experiments. The upper part of the tank is filled with water with a linear temperature profile from $4\degree {\rm C}$ to $22\degree {\rm C}$, using the double-bucket technique frequently used in salted-water experiments \cite[see, e.g.,][]{Oster65}. Note that this filling process allows us to avoid a very long transient (typically 1 day) before reaching a quasi-steady state set by thermal diffusion in the upper stratified zone. The system is typically allowed to evolve for 1 hour before measurements are taken.

We have performed two types of measurements. First, two high-precision Platinium resistance sensors with diameter $1 \ \text{mm}$ and acquisition frequency $2 \ \text{Hz}$ give access to temperature fluctuations at chosen locations with a precision of $\pm 0.001\degree {\rm C}$ \cite[more than one order of magnitude better than the preliminary measurements presented in][]{perrard}. These two probes are mounted on a vertical linear stage with a vertical spacing of $12 \ \text{cm}$, positioned in the middle of the tank in the horizontal directions. Vertical time-averaged temperature profiles, such as the one shown in figure \ref{fig_setup}($b$), are determined by taking measurements at regularly spaced locations. The measured profiles are characteristic of a two-layer system, with a monotonically increasing temperature in the upper stratified zone, and a fully mixed lower convective zone of constant temperature at about $3.4\degree {\rm C}$ in the horizontal middle of the tank, i.e. where rising cold structures are expected (see figure \ref{fig_setup}$c$). 
The observed depth of the convective zone is $h_{\rm conv} = 20 \ \text{cm}$ for the well-insulated temperature measurements: this corresponds to a typical Rayleigh number $Ra$ of about $5\times 10^7$. 
For the less-insulated experiments involving velocity measurements, $h_{\rm conv} = 9 \ \text{cm}$, corresponding to a smaller Rayleigh number $Ra=5\times 10^6$. In both cases however, $Ra$ is above the expected transition for turbulent behavior \cite[][]{Krishnamurti81}. 
The temperature gradient in the upper stratified zone is not rigorously constant because of side heat losses, despite efforts to minimize them. Combined with the peculiar equation of state of water, this leads to a non-constant buoyancy frequency in the stratified zone, $N$ \cite[see also][]{townsend1964natural, perrard, Lecoanet2015}. However, to simplify our analysis, we approximate $N$ to be constant and equal to its mean value in the $6.5 \ \text{cm}$ region extending above the $8\degree {\rm C}$ interface, where we have measured propagating waves in our set-up. This value is $N \simeq 0.06 \ \text{Hz}$ when using full insulation, and $ N \simeq 0.07 \ \text{Hz}$ otherwise. Once these mean properties of the flow are established, long measurements of temperature fluctuations with time at selected locations are recorded over 3 days, and analyzed in detail in section 3.

In addition to temperature measurements, we also perform velocity measurements using particle image velocimetry (PIV). The water is then seeded with spherical reflective particles of typical diameter $10 \ \mu\text{m}$. The tank is illuminated from one of the short sides using a $40 \ \text{mW}$ green laser vertical light sheet, and a camera records the particles displacements from one of the long sides at $4 \ \text{fps}$ with a resolution of $1920 \times 1080$ pixels for the convective zone, and $0.5 \  \text{fps}$ with a resolution of $3680 \times 2456$ pixels for the stratified zone. The captured movies of typical duration $10 \ \text{min}$ are PIV-processed using the freely available software DPIVSoft \cite[][]{Meunier03}. The velocity fields are typically resolved into $32\times 32$ pixels boxes with $50\%$ of overlapping, leading to a spatial resolution of about $1.3 \ \text{mm}$ in the stratified zone. Figure \ref{fig_setup}($c$) shows a sketch of the expected velocity profile, with a large-scale circulation plus fluctuations in the convective zone and internal waves in the stratified zone, as will be demonstrated in section 4. 
\\

%%%%%%%%%%%%%%%%%%%%%%%%%%%%%%%%%%%%%%%%%%%%%%%%%%
%%%%%%%%%%%%%%%%%%%%%%%%%%%%%%%%%%%%%%%%%%%%%%%%%%
\section{Analysis of the temperature signals} \label{sectionTemp}

\begin{center}
\begin{figure}
\begin{center}\includegraphics[width=\linewidth]{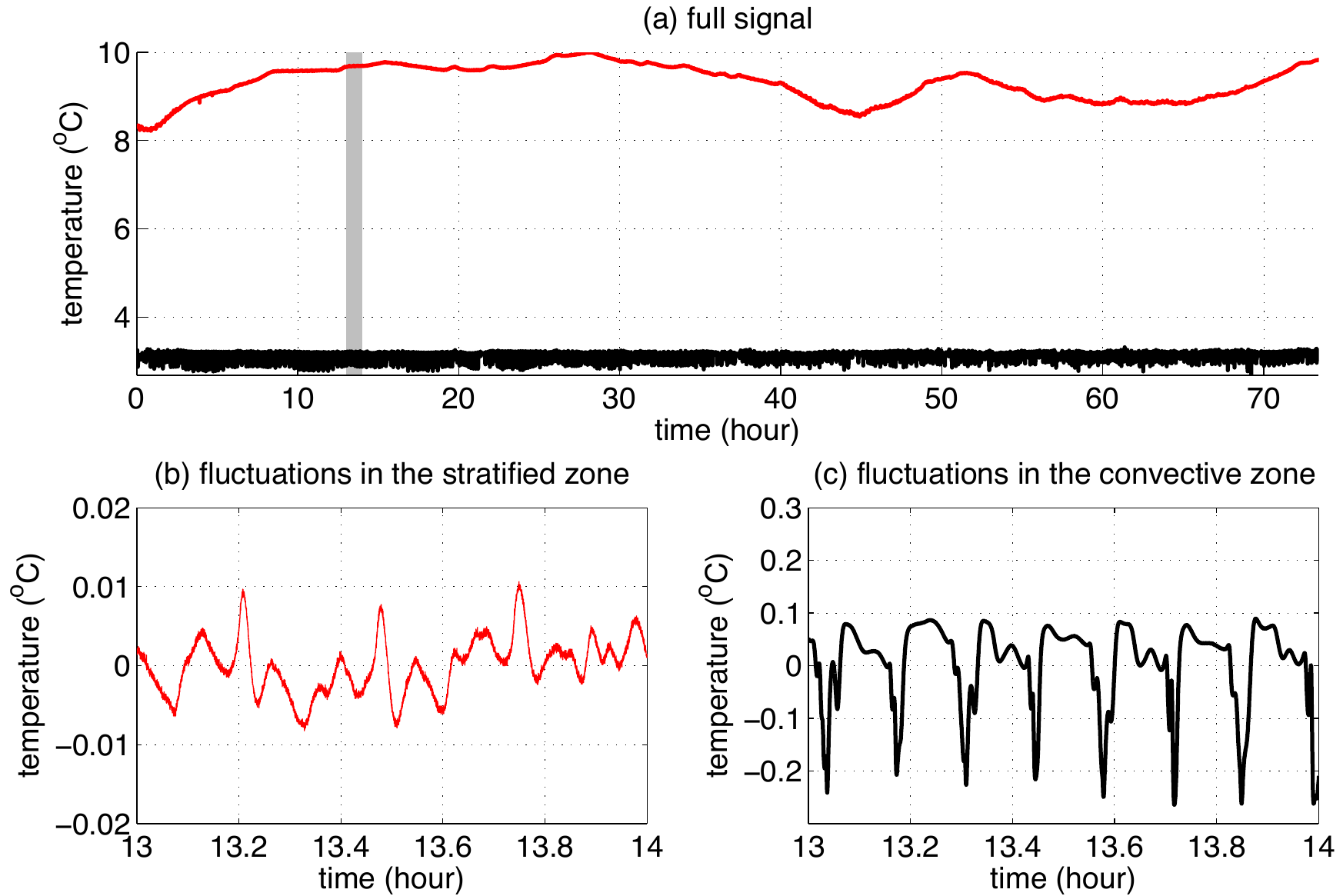}\end{center}
\caption{($a$) Temperature measurements during 3 days at fixed locations in the convective zone (lower black signal) and in the stratified zone (upper red signal) as a function of time in hours. ($b$) and ($c$) show a zoom of the corresponding fluctuations between 13 and 14 hours---shown by the gray rectangle in ($a$)---once secular trends have been removed by a polynomial fit. }
\label{fig_data_T}
\end{figure}
\end{center}

Figure \ref{fig_data_T} shows temperature data recorded over 3 days at fixed locations in the convective and stratified zones, at respective heights of $10 \ {\rm cm}$ and $22 \ {\rm cm}$ from the bottom of the tank. Despite our efforts to minimize lateral heat losses, the temperature in the stratified zone is still very sensitive to the room temperature, whose variations of about $3\degree {\rm C}$ between day and night explain the secular variations observed in figure \ref{fig_data_T}($a$). Figures \ref{fig_data_T}($b$) and ($c$) show the temperature fluctuations over one hour after removing the secular variations, in the stratified and convective zones, respectively. Note the different temperature scales, with convective temperature fluctuations in panel ($c$) being more than one order of magnitude larger than internal wave temperature fluctuations in panel ($b$), even though the stratified probe is located only $\sim 2 \ \text{cm}$ above the interface. Temperature fluctuations in the convective zone shown in figure  \ref{fig_data_T}($c$) are clearly skewed, with sharp negative peaks corresponding to  cold plumes rising in the middle of the tank. On the contrary, temperature fluctuations in the stratified zone shown in figure  \ref{fig_data_T}($b$) are more symmetric, as expected from  temperature fluctuations associated with propagating internal waves. 

\begin{center}
\begin{figure}
\begin{center}\includegraphics[width=\linewidth]{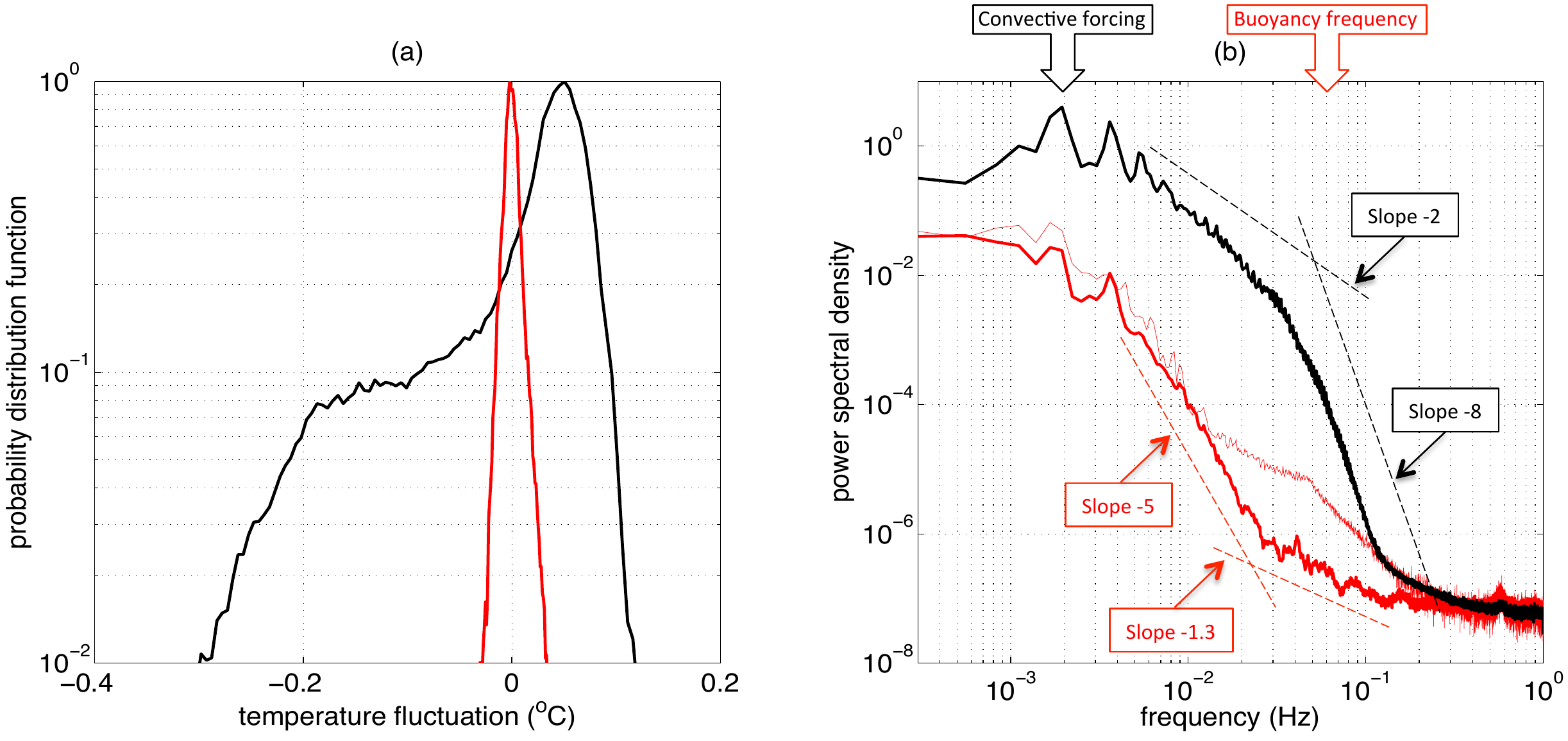}\end{center}
\caption{($a$) Normalized probability density function and ($b$) power spectral density (obtained with the Matlab function {\it pwelch}) of the temperature signals shown in figure \ref{fig_data_T}($a$), the data in black corresponding to the temperature fluctuations in the convective zone and the data in red corresponding to the temperature fluctuations in the stratified  zone. In ($b$), the thin red line shows the power spectral density obtained from the first 8 hours of measurements, while the thick  red line shows the power spectral density obtained from the following 64 hours of measurements. Typical power law dependences are indicated as dashed lines. The slopes corresponding to the convective signal  are estimated from the data published in \cite{wu90}. The slopes corresponding to the stratified signal are indicative only.}
\label{fig_spectral_analysis_T}
\end{figure}
\end{center}

This conclusion is confirmed in figure \ref{fig_spectral_analysis_T}($a$), which shows the probability density function of each of the measured signals normalised by its maximum value. To account for the longterm variations shown in figure \ref{fig_data_T}($a$), the raw temperature measurements are cut into 4-hours long signals, from which secular variations are removed by a polynomial fit of degree 10 in the stratified zone and degree 1 in the convective zone. Each result shown here then corresponds to the average probability density function over the 18 signals obtained from the 3-days measurements. 
The statistics of the temperature fluctuations in the convective region is comprised of two superimposed Gaussians: one from the convective turbulence, and the other from the ascending cold plumes. On the contrary, the statistics of the temperature fluctuations in the stratified zone shows an exponential distribution, which is due to the intermittent generation of internal waves.

We also computed the power spectral densities (PSD) of the same signals, which are shown in figure \ref{fig_spectral_analysis_T}($b$). The upper black convective signal is typical of turbulent Rayleigh-B\'enard convection, with a decreasing negative slope towards high frequency. The peak at $2~\times~10^{-3} \ \text{Hz}$ corresponds to the overturning timescale, and is in good agreement with the observed period of temperature fluctuations (see figure \ref{fig_data_T}c). We show for comparison the power laws that we estimated from the data published in \cite{wu90}, obtained in a classical Rayleigh-B\'enard experiment at Rayleigh numbers $7\times 10^6~-~1.1\times 10^8 $. The good agreement with our measurements suggests that the presence of the stratified layer does not significantly modify the behavior of the convective layer.

As expected, the power spectral density in the stratified zone is much smaller than in the convective zone (figure \ref{fig_spectral_analysis_T}$b$). We noticed a large difference in the stratified zone PSD at intermediate frequencies between the first 8 hours of measurements (thin red line) and the following 64 hours (thick red line).  But in both cases, no favorite frequency of wave propagation appears in the data analysis. We discuss the similar features of the two PSD curves below.

Integrating the signals between frequencies $0$ and $N$, we estimate the ratio between the wave energy and the convective energy to be about $1\%$ at this depth, consistent with the estimates of \cite{ansong2010internal}. The broad features of the stratified PSD follow those of the convective PSD, with most of the energy concentrated at low frequency and decreasing power with increasing frequency. However, there is a clear change of behavior around the frequency $2 \times 10^{-2} \ \text{Hz}$. From the main convective excitation frequency at $2 \times 10^{-3} \ \text{Hz}$ to this frequency $2 \times 10^{-2} \ \text{Hz}$, the PSD slope in the stratified zone is significantly steeper than the slope in the convective zone. This is in qualitative agreement with the analytical model of \cite{lecoanet2013internal}, which studied the waves produced by Reynolds stresses from Kolmogorov-type turbulence. As described by \cite{lecoanet2013internal}, the underlying physical explanation is that large frequency excitation sources a priori correspond to smaller lengthscale fluctuation patterns: their e-folding depth of evanescence during their propagation in the convective zone is thus smaller. Additionally, one also expects smaller structures to be more rapidly damped by viscous and thermal diffusion, both in the convective and stratified zones. On the contrary, above $2 \times 10^{-2} \ \text{Hz}$, the PSD slope in the stratified zone is significantly less steep than the slope in the convective zone. We argue that this is directly related to the existence of propagating internal waves, and to their peculiar property that for a given horizontal wavelength, the damping length is an increasing function of frequency up to $\omega = 2N/\sqrt{5}$, as already noticed by \cite{Taylor2007internal} in the case of waves generated by a turbulent bottom Ekman layer. 

As in \cite{michaelian2002coupling}, we see a large difference of two orders of magnitude between the early and the late temperature PSD.  We attribute this to more plume activity within the convection zone during the early times of the experiment, even if no sign of this randomness is seen in the statistics of the convective signals. 
Finally, the slope of the PSD of the first 8 hours of data (thin red line) steepens near the buoyancy frequency $N$.  This is consistent with the transition between propagative internal waves expected for frequencies below $N$, and evanescent internal waves expected for frequencies above $N$. This is not observed in the PSD of the following 64 hours measurements, probably because of the limited precision of the temperature measurements. The spectrum then converges towards white noise, which can also been seen in the convective spectrum. 
\\

%%%%%%%%%%%%%%%%%%%%%%%%%%%%%%%%%%%%%%%%%%%%%%%%%%
%%%%%%%%%%%%%%%%%%%%%%%%%%%%%%%%%%%%%%%%%%%%%%%%%%
\section{Analysis of the velocity signals} \label{secVel}

\begin{center}
\begin{figure}
\begin{center}\includegraphics[width=\linewidth]{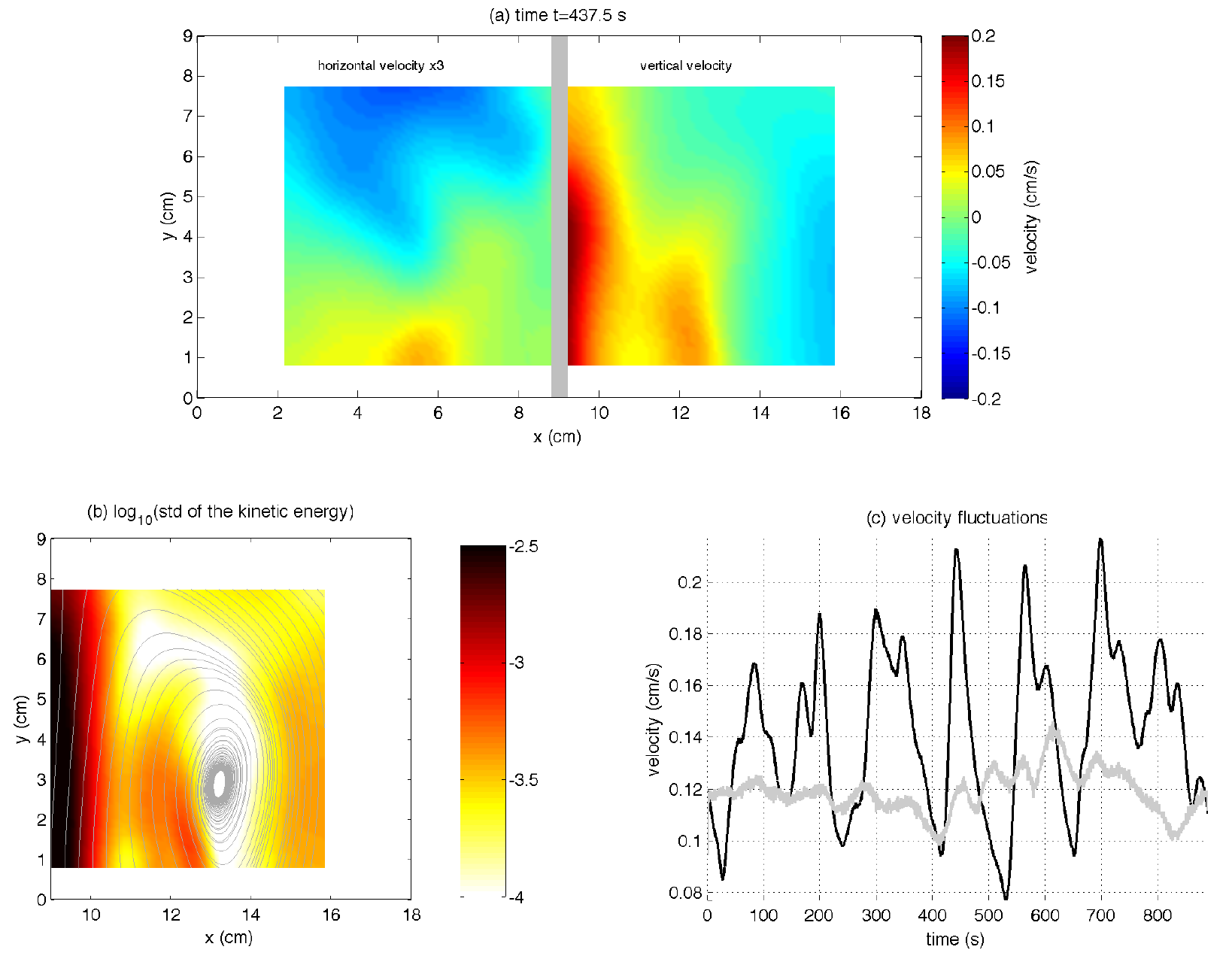}\end{center}
\caption{($a$) Example of the velocity field measured by PIV in the convective zone, with the color-plot on the left of the gray rectangle corresponding to three times the horizontal velocity and the color-plot on the right of the gray rectangle corresponding to the vertical velocity. Note that both signals come from the analysis of the same PIV movie taken on the right side of the tank.  We flipped the horizontal field across the central axis of the tank for visualization purposes. See also in supplementary material the corresponding movie ``ConvectiveField.avi'' showing the time evolution of those velocity measurements over $885 \ {\rm s}$. ($b$) Logarithm of the standard deviation of the velocity squared over $885 \ {\rm s}$ (in $({\rm cm}/{\rm s})^2$) and streamlines of the time-averaged velocity field, illustrating the mean organisation of the flow and the location of the main velocity fluctuations. ($c$) Example of the measured velocity fluctuations as a function of time: with reference to figure ($a$), the black line shows the vertical velocity at location $x=9 \ \text{cm},y=4.5 \ \text{cm},$ and the gray line shows the horizontal velocity multiplied by $-3$ at location $x=5 \ \text{cm},y=7.5 \ \text{cm}$.}
\label{fig_conv0}
\end{figure}
\end{center}

To perform velocity measurements in our set-up, the foam insulation is removed from the two long sides of the tank. As a result, the side heat losses significantly increase, and the typical depth of the convective zone is now $9 \ \text{cm}$. Figure \ref{fig_conv0} presents the main characteristics of the velocity field in the convective zone (see the corresponding movie ``ConvectiveField.avi'' in the supplementary material). As illustrated in figure \ref{fig_conv0}($a$), the velocity field consists mostly of two symmetric large cells filling the whole depth of the convective layer, with strong, localized, rising cold plumes in the center and broader sinking motions on the sides. Figure \ref{fig_conv0}($b$) shows the streamlines interpolated from the mean 2D flow determined by averaging the velocity measurements for $885 \ {\rm s}$, starting from initial points regularly spaced in the x-direction on the horizontal lines $y=0.8 \ {\rm cm}$ and $y=7.7 \ {\rm cm}$ (i.e. the bottom and top of the PIV domain). This figure also shows the associated standard deviation of the kinetic energy, as a proxy for quantifying the velocity fluctuations superimposed on the mean flow. Velocity fluctuations are mostly localized along the central axis of the tank, related to the passage of cold plumes forming along the lower boundary and then advected by the mean flow. The temporal evolution of the vertical and horizontal velocities at the locations where they are maximal is shown in figure \ref{fig_conv0}($c$). A typical period of about $75 \ \text{s}$ dominates both signals, corresponding to a frequency of about $0.013 \ \text{Hz}$. This main convective excitation frequency is significantly larger than the main convective excitation frequency of $2~\times~10^{-3} \ \text{Hz}$ observed in the temperature measurement experiment in section \ref{sectionTemp}. This is probably due to the fact that the height of the convective layer is significantly smaller in the present case, while the injected heat flux is the same in both cases, corresponding to a temperature drop of about $4^oC$ through the lower conductive thermal boundary layer of about $0.6  \ \text{cm}$ \cite[see ][]{perrard}: we thus expect the turn-over timescale to be shorter for the smaller convective cells observed here.

\begin{center}
\begin{figure}
\begin{center}\includegraphics[width=.9\linewidth]{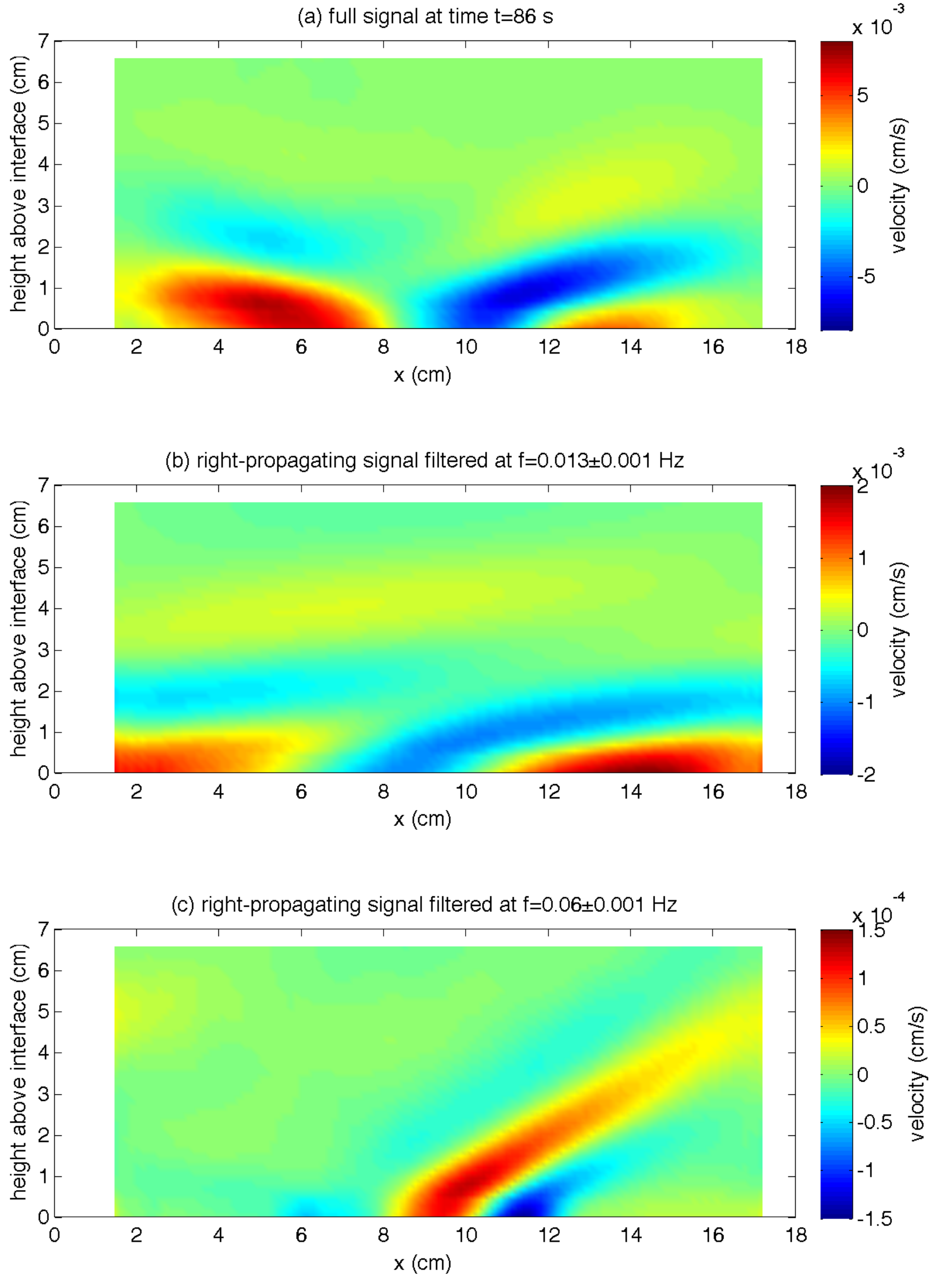}\end{center}
\caption{($a$) Example of the horizontal velocity field measured by PIV in the stratified zone. ($b$) and ($c$) show the corresponding signals including right-propagating waves only, respectively filtered around the frequency $f=0.013 \ {\rm Hz}$ (i.e., close to the typical convective frequency) and around the frequency $f=0.06 \ {\rm Hz}$ (i.e., close to, but below, the buoyancy frequency). Note the changes in the color scale, with interfacial velocities in panel ($b$) being more than one order of magnitude larger than interfacial velocities in panel ($c$). See also in the supplementary material the corresponding movie ``InternalWaves.avi'' showing the time evolution of these velocity measurements over $300 \ {\rm s}$. }
\label{fig_ondes}
\end{figure}
\end{center}

\begin{center}
\begin{figure}
\begin{center}\includegraphics[width=.9\linewidth]{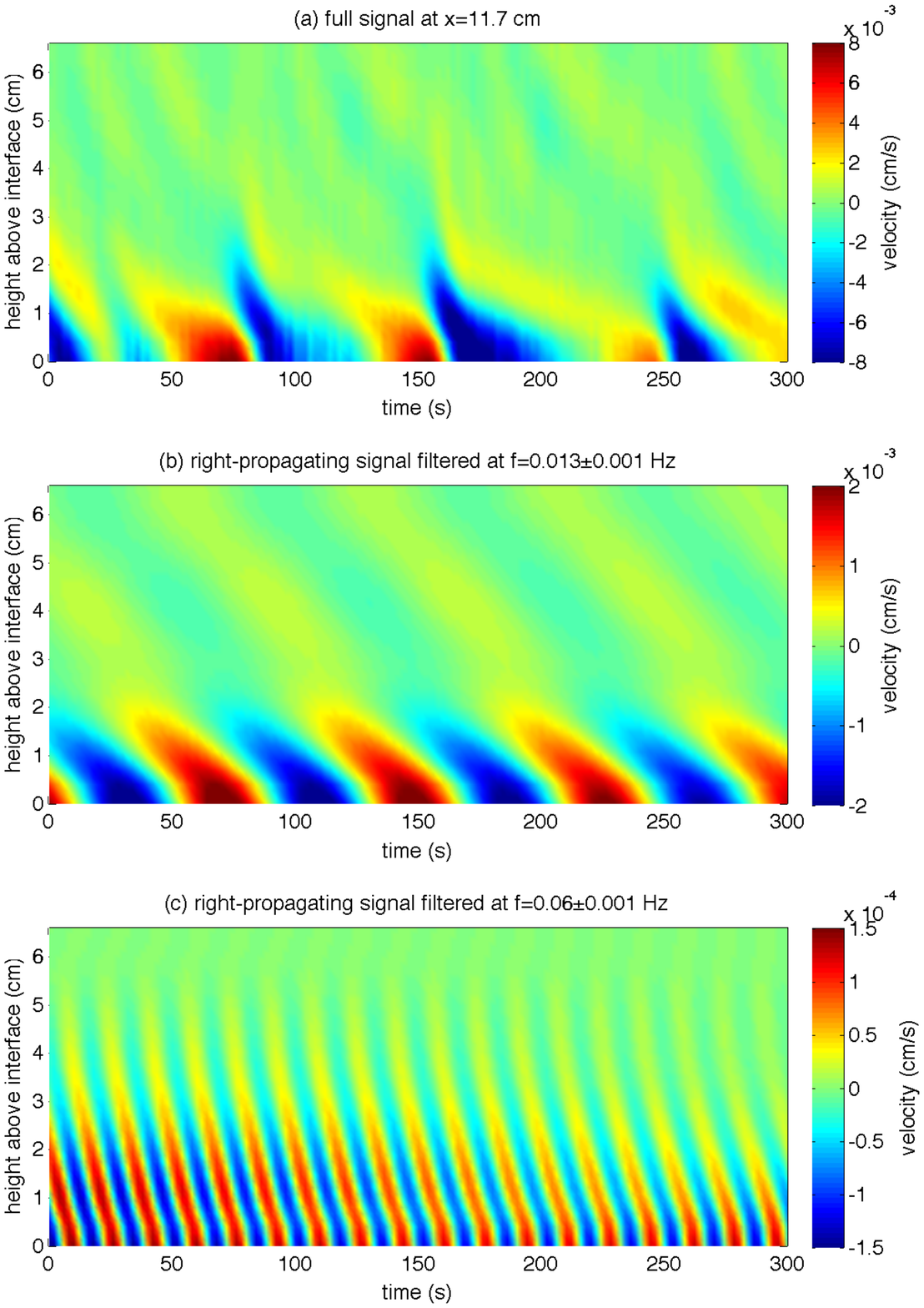}\end{center}
\caption{Space time diagrams showing the time evolution along the vertical line at $x=11.7 \ \text{cm}$ of the horizontal velocity field measured by PIV in the stratified zone (see the corresponding movie ``InternalWaves.avi'' in the supplementary material). As in figure \ref{fig_ondes}, ($a$) shows the full signal, while ($b$) and ($c$) show the corresponding signals including right-propagating waves only, filtered around the frequency $f=0.013 \ {\rm Hz}$ and $f=0.06 \ {\rm Hz}$, respectively.}
\label{fig_espacetemps}
\end{figure}
\end{center}

The temporal evolution of the wave field excited by this convective flow is measured by PIV during sequences of more than 10 minutes. A movie of measured wave velocities is provided in the supplementary materials (``InternalWaves.avi''). 
Figure \ref{fig_ondes} shows a frame of this movie, and figure \ref{fig_espacetemps} shows the corresponding space-time diagram obtained by extracting data from the vertical line at $x=11.7 \ \text{cm}$, so as to illustrate the wavy character of the velocity pattern. Figures \ref{fig_ondes}($a$) and \ref{fig_espacetemps}($a$) show the complete signal. Measured wave velocities are typically 25 times smaller than the convective velocities, and decrease rapidly away from the interface. The flow shown here corresponds to the superposition of waves traveling in all directions and with a wide range of frequencies. To better illustrate the wave patterns, we first spatially filter the complete velocity signal to show only right-propagating waves. In our case the energy goes from the interface towards the top of the tank, so these correspond to wave vectors oriented towards the bottom right. We also filter our signal in time to only show waves in a narrow frequency band of $\pm0.001 \ \text{Hz}$ around a mean filtering frequency. In figures \ref{fig_ondes}($b$) and \ref{fig_espacetemps}($b$), we chose the filtering frequency $f=0.013 \ \text{Hz}$, i.e. the typical frequency from the convective field (see e.g. figure \ref{fig_conv0}c). In figures \ref{fig_ondes}($c$) and \ref{fig_espacetemps}($c$), we chose the filtering frequency $f=0.06 \ \text{Hz}$, close to but below the mean buoyancy frequency of the stratified zone $N \simeq 0.07 \ \text{Hz}$.

The two types of waves previously described in the transient experiment of \cite{michaelian2002coupling} are clearly recovered in our data. Most of the energy is in waves with long horizontal wavelength and low frequency as shown in figures \ref{fig_ondes}($b$) and \ref{fig_espacetemps}($b$). Those waves concentrate close to the interface, and are rapidly damped as they propagate away from it. This region close to the interface corresponds to the buffer zone that was described by \cite{perrard}, where the buoyancy frequency rapidly changes from zero in the convective zone to a constant positive value, and where low frequency waves  propagate at angles as small as $10\degree$. Additionally, some energy is carried by waves with shorter horizontal wavelengths and higher frequencies, as shown in figures \ref{fig_ondes}($c$) and \ref{fig_espacetemps}($c$).

\begin{center}
\begin{figure}
\begin{center}\includegraphics[width=.8 \linewidth]{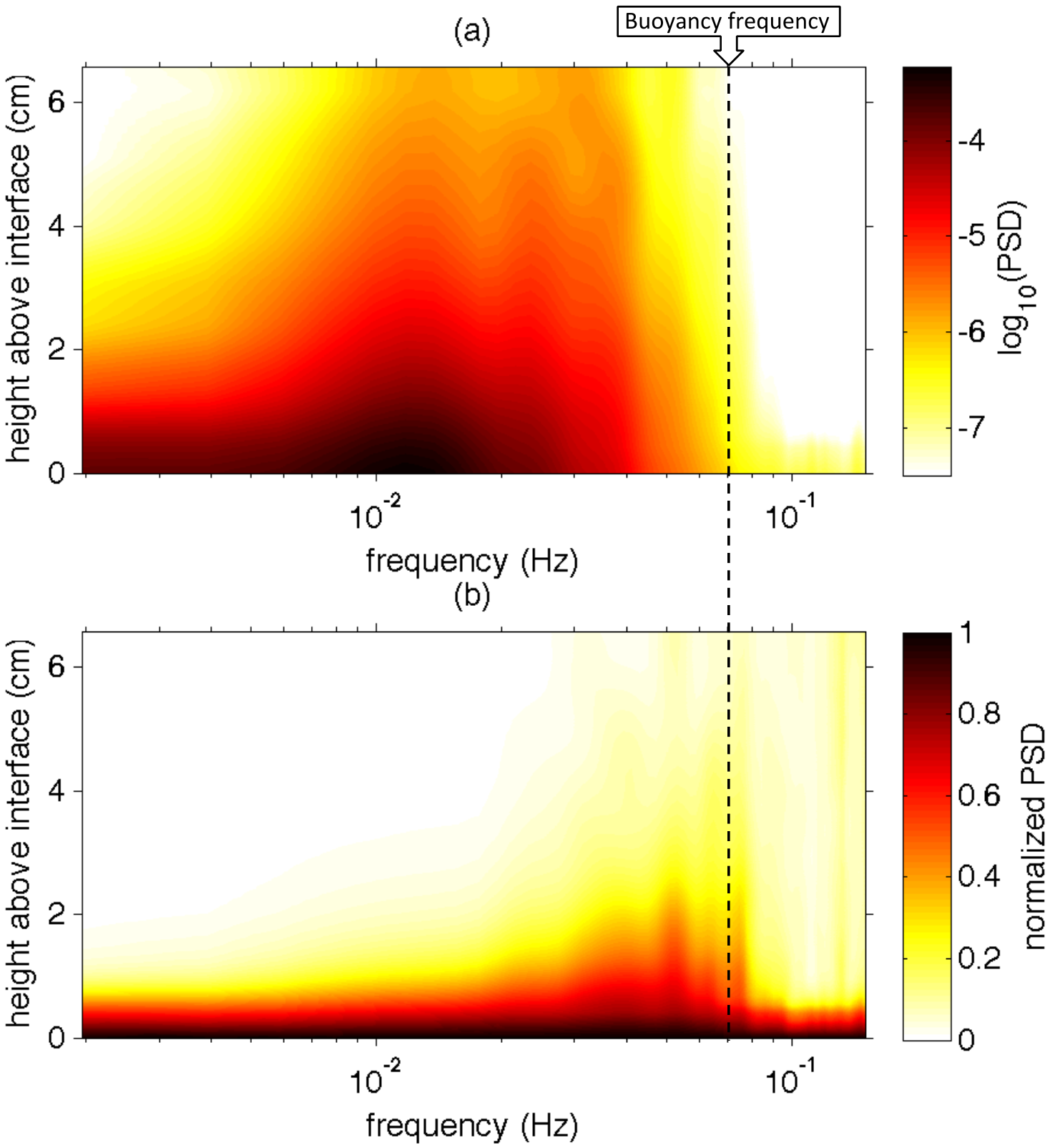}\end{center}
\caption{Spectrograms of the horizontal velocity in the stratified zone, measured by PIV over 13 minutes (see an extract of the data in figures \ref{fig_ondes} and \ref{fig_espacetemps} as well as in the movie ``InternalWaves.avi'' in the supplementary materials). Along each horizontal line in ($a$), the color plot shows the logarithm of the $x$-averaged power spectral density at a given depth, obtained using the Matlab function {\it pwelch}. ($b$) shows the same results when renormalized by the amplitude of the power spectral density at the interface ($y=0$).  }
\label{fig_spectrogram}
\end{figure}
\end{center}

To further prove this last point, we show in figure \ref{fig_spectrogram} spectrograms of the velocity signal as a function of height above the interface, with ($a$) showing the raw result and ($b$) showing the result normalized by the power spectral density at the interface. In both cases, a cut-off frequency close to the typical buoyancy frequency $N \simeq 0.07 \ \text{Hz}$ appears: energy is carried in the stratified zone by propagating waves only. Also, it clearly appears from the renormalized spectrogram in ($b$) that excited waves with frequency close to $N$ are less rapidly damped, confirming our interpretation of the temperature spectrum in section \ref{sectionTemp}. No favorite frequency of propagation appears here: the observed spectrum results from a competition between the way the injected wave energy is distributed
among the different horizontal wavelengths and temporal frequencies, and the way the waves are damped away from the interface. 
\\

%%%%%%%%%%%%%%%%%%%%%%%%%%%%%%%%%%%%%%%%%%%%%%%%%%
%%%%%%%%%%%%%%%%%%%%%%%%%%%%%%%%%%%%%%%%%%%%%%%%%%
\section{Conclusion}
We have re-investigated the ``$4\degree {\rm C}$ experiment'' first introduced by \cite{townsend1964natural}, which allows a single self-organising system to produce a  turbulent convective layer adjacent to a stably stratified one, as is encountered in atmospheric and stellar flows. 
We provide here a set of quantitative high-precision measurements that can help validate analytical and numerical models of such complex systems, whose dynamics range over wide time, length and velocity scales. The combination of spectral analysis of the temperature fluctuations and PIV velocity measurements allows us to describe the complex internal wave field excited in the stratified zone by the turbulence. In agreement with previous studies of closely-related configurations \cite[][]{michaelian2002coupling, ansong2010internal}, we describe the flow pattern as a superimposition of low frequency, long horizontal wavelength waves close to the interface, and higher frequency, smaller horizontal wavelength waves propagating away from the interface.

Previous studies also report the existence of a favorite angle of propagation, which was explained by \cite{dohan2003internal} as a resonant coupling between turbulent fluctuations and the excited internal waves that optimize the vertical transport of horizontal momentum. Such a selection is not observed in our experiment, where waves are of very small amplitudes: this conclusion supports a non-linear origin of the angle selection process in previous experiments, where larger amplitude waves were excited. In the present case, we rather hypothesize that the internal wave field is passively excited by the convective field, and 
that the observed wave field results from two contributing physical mechanisms: (1) the spatiotemporal characteristics of the turbulent excitation; and, (2) the selective damping of small wavelength waves and/or of low frequency waves during their propagation in the stratified zone, as fully described by the dissipative, linear, plane wave theory \cite[][]{Taylor2007internal, Lecoanet2015}.  
The remaining question of the main process of wave excitation, i.e. interface deflections or bulk Reynolds stresses within the convective zone, is addressed in the paper \cite{Lecoanet2015}, which studies direct numerical simulations and simplified numerical models of our experiment. Non-linear effects potentially responsible for zonal flow generation and for frequency selection will be addressed in future work, involving a larger, three-dimensional, experimental set-up.
\\

%%%%%%%%%%%%%%%%%%%%%%%%%%%%%%%%%%%%%%%%%%%%%%%%%%
%%%%%%%%%%%%%%%%%%%%%%%%%%%%%%%%%%%%%%%%%%%%%%%%%%
M.L.B. acknowledges support from the Marie Curie Actions of the European Commission (FP7-PEOPLE-2011-IOF). M.L.B. and D.L. thank the Labex program MEC (ANR-11-LABX-0092) for supporting D.L. visit in Marseilles.

\bibliography{biblio}

\end{document}